\documentclass[%
showkeys,
preprint,
amsmath,amssymb,
aps,
pra,
]{revtex4-1}
\usepackage{graphics}
\usepackage{epsfig}
\usepackage{epstopdf}
\usepackage{amsmath}
\usepackage{array}
\usepackage{subfigure}
\usepackage[colorlinks=true,linkcolor=blue,citecolor=blue,      urlcolor=blue]{hyperref}% add hypertext capabilities
\begin{document}
	\title{Emergence of space from non-equilibrium thermodynamics in f(R) gravity}
%\title{Emergence of space from the generalized first law of thermodynamics in f(R) gravity for (n+1) FRW Universe }
%	\title{Emergence of space in f(R) gravity for n dimensional FRW Universe and thermodynamics }
    %\title{Modified expansion law in f(R) gravity and thermodynamics }
	\author{Hassan Basari V. T.}
	\email{basari@cusat.ac.in}
	\author{P. B. Krishna}
	\email{krishnapb@cusat.ac.in}
	\author{Titus K. Mathew.}
	\email{titus@cusat.ac.in}
	\affiliation{%
		Department of Physics, Cochin University of Science and Technology, Kochi, Kerala 682022, India
		}%
\begin{center}
%{\bf \Large {A Cosmological Model from Emergent Gravity Paradigm}} \\[0.2in]
%(titus@cusat.ac.in)
%\\[0.3in]
\begin{abstract}
 It has shown that the accelerated expansion of the FRW Universe can be explained as the quest towards the holographic equipartition ($N_{sur} = N_{bulk}$), satisfies the expansion law $\frac{dV}{dt} = l_{P}^{2} \left( N_{sur} - \epsilon N_{bulk} \right)$, from which one can derive the Friedmann equation of the FRW Universe in Einstein gravity \cite{paddy2012jun}. We introduce a generic derivation of the expansion law from the generalized first law of thermodynamics $-dE=TdS$ and $dE=TdS + WdV$. The generic derivation provides an
 % generic 
 expression for $N_{sur}$ in terms of entropy $S$ and the 
 %generic 
 expansion law consistent with gravity theories having different entropy $S$, like Gauss-Bonnet and more general Lovelock gravity. We extended the same idea to the non-equilibrium situation and obtained the expansion law in f(R) gravity as a specific case. For this, we used the first law of thermodynamics in non-equilibrium description having the extra entropy production term $Td_{i}S.$ 
\end{abstract}  
\maketitle
\end{center}
\section{Introduction}
%The Thermodynamic aspects of gravity were explored first from the studies in black hole physics. 
Historically, the connection between gravity and thermodynamics emerges from studies on black hole physics. In 1973, Bekenstein introduced the concept of black hole entropy as a measure of the horizon area of the black hole \cite{bekenstein1}. Using this, the same author generalised the second law of thermodynamics for a region containing
a massive black hole \cite{bekenstein2}. Parallelly Bardeen et al. formulated the four laws of black hole mechanics, which are analogous to the laws of thermodynamics of an ordinary macroscopic system \cite{bardeen}. During the same time, Hawking \cite{Hawking1} showed that the black holes, formed by the collapse of matter, can radiate particles with a thermal spectrum having temperature $T=\kappa/2\pi,$ where '$\kappa$'  is the surface gravity at the horizon. The temperature is not a peculiar feature of the black hole alone but a general feature of all spacetime horizons \cite{PhysRevD.7.2850, Davies_1975,PhysRevD.15.2738}. Remarkably,  
an accelerating observer in a flat spacetime, who possesses a horizon, 
can attribute a temperature proportional to the acceleration $ `a`$ as $ T=a/2\pi$ (Unruh temperature)
\cite{PhysRevD.14.870}. All these imply that Einstein's field equations of gravity on the spacetime horizon possess a thermodynamic structure.

Jacobson \cite{1995} showed that 
Einstein's field equations can be derived by projecting the Clausius relation, $\delta Q = T dS$ 
to a local Rindler horizon, where the horizon entropy is proportional to horizon area as proposed by Bekenstein \cite{bekenstein1}. Here $\delta Q$ is the energy flux through the local
Rindler horizon and $T$ is the Unruh temperature \cite{PhysRevD.14.870} seen by the accelerated observer near the horizon. Later, Cai and Kim \cite{Cai_Kim} derived the Friedmann equations by applying the first law of the form $-dE = TdS $  
to the apparent horizon of an (n+1)-dimensional FRW Universe. 
Here 
$S= A/4G$ is the horizon entropy following the Bekenstein formula,
$T=\dfrac{1}{2\pi r_{A}}$ is the temperature of the horizon  
and  $ -dE = A(\rho+p)Hr_{A}dt $ is the energy flux through the apparent horizon of radius $r_{A}$.
These authors extended 
this result in Gauss-Bonnet gravity and the more general Lovelock gravity by adopting the 
corresponding entropy.

Padmanabhan has shown that %proposed an alternative formulation for the thermodynamic law, 
the structure of the Einstein equation 
%tensor 
near any spherically symmetric horizon 
%has can be of 
assumes the form $dE= TdS -PdV,$ where $E$ is the energy associated with the horizon, which is proportional to the horizon's radius, $P$ here is the pressure term due to the gravitational source (which is different form $p$ used before) and $T$ is horizon temperature proportional to surface gravity  \cite{Padmanabhan:2002sha, PhysRevD.74.104015}. Later, Akbar and Cai derived the Friedmann equations by applying the first law of thermodynamics of the form,  $ dE = TdS + WdV$ \cite{PhysRevD.75.084003}, often called as the unified first law, to the apparent horizon of FRW Universe.  
For this unified first law, 
$ dE $ here is the distortion in the Misner-Sharp energy of the matter inside the apparent horizon, temperature $ T=\kappa /2\pi, $ a measure of the surface gravity as said before, and $ W =-\dfrac{1}{2}T^{ab}h_{ab} = \dfrac{1}{2}(\rho - p)$ is the work density, the orthogonal projection of energy-momentum tensor $T^{\mu \nu}$ to the horizon. Interestingly this form of the first laws found to have the same structure as 
the first law for the trapping horizons proposed by Hayward \cite{Hayward_1998,Hayward:1998ee}. It should be noted that for spherically symmetric horizon the work density 
%the apparent horizon, the pressure term $-p$ in the first law of the form, $dE= TdS -pdV$ became the work density 
$W$, is the effective pressure as defined by Padmanabhan, i.e. $ W= -(1/2)T_{ab}h^{ab}=-P, $ in the relation $dE= TdS -PdV$ \cite{PhysRevD.83.024026}.    
%Hence both expressions have the same structure at the apparent horizon. Notably, now we have two kinds of the generalized first law of thermodynamics; both have different aspects and structure \cite{PhysRevD.83.024026}.    

The 
%arguments and 
results shown above have clearly indicated that gravity is intimately connected to the thermodynamics of spacetime.
The thermodynamics of any ordinary  
%macroscopic 
system, say like a gas, is described 
%with its 
using the macroscopic variables like pressure,
temperature, etc., having no significance in the microscopic domain.  Hence these macroscopic variables 
of a system are considered to be 'emerged' from the collective behaviour of the
constituent microscopic degrees of freedom associated with the system. Similarly, the connection between gravity and thermodynamics implies that 
%gravity with its
macroscopic properties like 
%including 
metric, curvature, etc. are irrelevant in 
%going over to 
the microscopic
domain, but could be emerged in the macroscopic domain due to collective behaviour of some underlying fundamental microscopic degrees of freedom.  These motivate  
the reformulation of gravity as an emergent phenomenon.
In this context, Padmanabhan
\cite{paddy2010dec} derived Newton's law of gravity by combining the equipartition law of energy for the degrees of freedom at the horizon and the
thermodynamic relation for entropy, $S = E/2T,$ where  
$E$ is the effective gravitational mass, and $T$ is the
horizon temperature. On the other hand, following string theory considerations, Verlinde 
%also 
reformulated gravity as an entropic force arising from the natural tendency of material distribution to maximize the entropy \cite{Verlinde}. These works %concludes 
enlightened the concept that gravity could be an emergent property in a pre-existing spacetime.

On further extending this idea of emergent paradigm, Padmanabhan has shown that, 
%the time evolution of the spacetime 
%In a general, 
for an evolving spacetime, the rate of change of gravitational momentum is proportional to the difference in the 
%between the 
number of 
%bulk and boundary 
degrees of freedom between the horizon and the bulk within the horizon, i.e.  $N_{sur} - N_{bulk}.$ The evolution will come to a stop when $N_{sur}=N_{bulk}$ and is known as the holographic equipartition. So ultimately, it is the 
% This 
departure from the holographic
equipartition, which drives the time evolution of spacetime. 
%itself can be described as 
%proportional to 
%On this emergent approach, one can formulate gravitational field equation From a thermodynamic action \cite{PhysRevD.75.064004, Padmanabhan:2007},
%\begin{equation}
%A[l, \nabla ] = \int  \dfrac{d^{4}x}{l_{p}^{4}}\sqrt{-g} \left(l_{p}^{4}T_{ab}l^al^b +P^{ab}_{cd}\nabla_{a}l^c\nabla_{b}l^d\right) ,
%\end{equation}
%where  $l^{a}$ is the null vector, $T_{ab} $ is the energy momentum tensor and $ P^{ab}_{cd} = \frac{l_{p}^{2}}{8\pi}
%\left(\delta^{a}_{c}\delta^{b}_{d}-\delta^{b}_{c}\delta^{a}_{d}\right)$ is the entropy functional for any vector field in terms of a fourth-rank divergence-free tensor with the symmetries of the curvature tensor \cite{PhysRevD.75.064004}. On this formalism, the time evolution of space time geometry in a region bounded by the $ \sqrt{-g_{00}}= $ constant surface can be shown that \cite{Padmanabhan:2013}, 
%\begin{equation}\label{eq:emergent gravity}
%\frac{1}{8\pi}\int \limits_{\mathrm{V}}d^{3}x\sqrt{h}u_{a}g^{ij}\L_{\zeta} N^{a}_{ij} = \frac{1}{2} k_{B}
%$T_{avg} \left( N_{sur} - N_{bulk} \right)$\cite{Padmanabhan:2013}.  
%%\end{equation}
%Here $ N_{sur} $  is the degrees of freedom $(DoF)$ on the horizon surface and $N_{bulk}$ is the $DoF$ on the bulk enclosed by the horizon
%%is the $DoF$ for the equipartition of Komar energy to the 
%and 
%%average horizon temperature
%$T_{avg}$ is the average temperature of the horizon. 
In line with this, Padmanabhan took one step further to propose that 
%also explored that the 
space itself possess an emergent nature in the cosmological context \cite{paddy2012jun}. He states that the expansion of the Universe (expansion of the Hubble volume) can be explained as the emergence of space with the progress of cosmic time $t$. 
%He asserts that 
It is conceptually difficult in general to consider the time has 
%to have 
emerged from
any pre-geometric variables. However, this difficulty will not be surfaced in cosmology due
to the existence of proper time for the geodesic observers for whom cosmic background radiation appears to be
homogeneous and isotropic \cite{paddy2012jun}.
%The emergence of space happens to equalize the degrees of freedom $(DoF)$ on the horizon and $(DoF)$ in bulk enclosed by the horizon. 
It was proposed that the time evolution of the Universe in Einstein's gravity can be described using the equation, $\dfrac{dV}{dt}= l_{p}^{2} \left( N_{sur} -\epsilon N_{bulk} \right),$ known as the holographic equipartition principle and later many called it as expansion
law, where $ V $ is the volume of the apparent horizon of the Universe and $l_p$ is the Planck length. The emergence of space happens to equalize the degrees of freedom $(DoF)$ on the horizon with the  $DoF$ in bulk enclosed by the horizon. Based on this paradigm, Padmanabhan derived the Friedmann equation from the expansion law for 
a flat FRW Universe in (3+1) Einstein gravity \cite{paddy2012jun}. 
%The idea of the emergence of space also offers a physical principle to measure the cosmological constant
%\cite{Paddy:2017intgratingconstant,paddycosmologicalconstant,paddyhamsacosmin}. 
The 
%above 
expansion law was 
%generalized and 
extended to higher dimensional gravity theories like (n+1) dimensional Einstein gravity, Gauss-Bonnet gravity and more general Lovelock gravity \cite{cai} by appropriately
modifying the surface degrees of freedom on the boundary surface.  An extension of this procedure to non-flat FRW Universe was done by Sheykhi\cite{Sheykhi2013}. More investigations in this line employing Padmanabhan's idea of emergent paradigm can
be found in references \cite{Padmanabhan2019review,paddycosmologicalconstant, PhysRevD.90.124017,SHEYKHI2018118,TU2018411,PhysRevD.99.043523,PhysRevD.88.084019,Tu_2013,FARAGALI, Yuan:2016pkz}. 
%20, 21, 22, 23, 24, 25 both in 
%for flat and non-flat FRW universe \cite{cai,Sheykhi2013}. For non-flat universe, the apparent horizon is considered in place of the Hubble horizon. 
%Even tough the above generalizations are very successful, some alternatives and criticisms about the generalizations also appeared in the literature \cite{FARAGALI}. Which seeks critical explanations for those generalizations other than the ansatz used by the authors \cite{cai,Sheykhi2013}.
In recent studies, one of us 
%already 
has shown that the expansion law effectively implies the entropy maximization in Einstein's gravity \cite{krishna1} and more general forms of gravity like Gauss-Bonnet and Lovelock gravities \cite{krishna2}.
Consequently, one can interpret the emergence of space in the expansion law as equivalent to the rise in horizon entropy with the progress of cosmic time and the Holographic equipartition, $N_{sur} =N_{bulk} $ as the state in which the horizon entropy is in its upper bound given in the references \cite{Fischler:1998st, Bousso_1999}.

In reference \cite{FLDezaki}, the authors  conjectured that the first law of thermodynamics could be considered the origin of the expansion law. They derived the expansion law using an appropriate form of the unified first law in (3+1) Einstein's gravity. In different approaches, the authors also used the same idea for Horava-Lifshitz gravity and $f(R)$ gravity. However, they failed to extend the idea to gravity theories like Gauss-Bonnet and Lovelock gravities. Recently it has 
%also 
been shown 
%recently, 
that the
expansion law can be derived from the first law of thermodynamics in Einstein's gravity, Gauss-Bonnet and Lovelock gravities \cite{Mahith2018}. 
%However, in this 
During the derivation process, the above authors have used the Friedmann equation on the midway. But the more important point is that the law of thermodynamics used is basically the equilibrium version of it. Hence there arise two questions, (i) whether the expansion law can be derived from the thermodynamic law without explicitly using the Friedmann equation and (ii) how to obtain the expansion law from the non-equilibrium version of the thermodynamic law. The first point is important in its own way, since it has already known that expansion law otherwise implies the Friedmann equation. So deriving the expansion law using the Friedmann equation is seems to be a circular process. The second point is relevant with reference to gravity theories 
% to reach the law of expansion.  
%Hence a proper generic derivation of the expansion law solely from the first law of thermodynamics is not found in the literature. 
%In this work, we first derive the expansion law from the first law of thermodynamics for the equilibrium description  in a generic manner, without relying on the Friedmann equations. 
%Such a generic connection will give generic expressions for the surface and bulk $DoF$ in different gravity theories and solve the disparities in the literature of fixing of the surface and bulk $DoF$ in gravity theories like Gauss-Bonnet and Lovelock gravities \cite{cai,FARAGALI}. Once we made a generic derivation for the expansion law, 
%Following this, 
%it is important to 
%we investigate the generalisation of 
%method  
%what extent 
%horizon thermodynamics 
%the expansion law 
%can be 
%developed 
%more general spacetime theories 
%against quantum gravitational effects. A feasible way to proceed in this
%direction is to weigh up the effects of 
in which higher-order curvature corrections 
%to Einstein gravity 
have been taken into account. In the presence of higher-order curvature corrections,  
the entropy becomes a polynomial function 
%in terms 
of the Ricci scalar. This in turn requires a non-equilibrium treatment, equivalent to non-equilibrium thermodynamics \cite{PhysRevLett.96.121301}. In such a situation, one has to 
%use 
%It has been shown that, to derive the 
%The 
%corresponding field equation, one has to 
use a modified  
%is derived 
%from the 
entropy balance relation $ \delta Q/T = dS+ d_iS,$ (or equivalently $TdS+Td_iS = -dE$) %equivalent to a 
often known as the non-equilibrium Clausius relation (or the first law of thermodynamics), where $d_iS$ corresponds to an additional entropy production due to non-equilibrium evolution happening in the system. The corresponding form of the unified first law in non-equilibrium will be $dE=TdS+WdV+Td_iS.$ 
%In short, we aim to derive 
%Hence secondly, we derived 
%the expansion law in the non-equilibrium description from the corresponding thermodynamic law.
%In our derivations, we concentrate on the 
%above modified first law of thermodynamics for the non-equilibrium description in a generic formulation. The derivation gives a more generic form of the expansion law, that can be used for the gravity theories having the higher order curvature corrections. Then, we will consider 
The $f(R)$ gravity is a typical example of considering the higher curvature corrections \cite{Nojiri-Odintsov:2011, Nojiri-Odintsov-vko:2017, Sotiriou:2008rp}, in which the action is an arbitrary function of the curvature scalar $R.$ In the present work we first derive the expansion law from the equilibrium thermodynamic law without explicitly using the Friedmann equation. We then extend this process to obtain the expansion law from the non-equilibrium thermodynamic law.
% which holds the Wald entropy $S=Af^{\prime}(R)/4G$ \cite{}.

The paper is organized as follows. In section \ref{sec.2}, 
%\textbf{we review the expansion law and its generalizations, then} 
we derive the general expansion law from the first law of thermodynamics for equilibrium description in a generic approach in section \ref{sec.3}. we extend this approach to the non-equilibrium description. In section \ref{sec.4}, we find the expansion law in f(R) gravity. Finally, we conclude our work in section \ref{sec.5}.

%\section{First law of thermodynamics and the emergence of space - equilibrium description}\label{sec.3}
\section{Expansion law from First law of thermodynamics - equilibrium case.} \label{sec.2}
%The expansion law maximizes the horizon area (areal volume of cosmic space) with the progress of cosmic time, this is to achieve the holographic equipartition. The area of the horizon have the direct link to the horizon entropy through the entropy-area relation, then one can express the L.H.S of the expansion law, $\dfrac{dV}{dt}$ in terms of entropy evolution $\dfrac{dS}{dt}$. This is not a mathematical coincidence, but it is already shown that the expansion law effectively implies the entropy maximization in $\Lambda$CDM model \cite{krishna1} and in more general cosmological models \cite{krishna2}. Moreover, it is also shown that the expansion law could be derived starting from the first  law of thermodynamics in (n+1) dimensional Einstein gravity, Gauss-Bonnet gravity and more general Lovelock gravities. Which altogether implies a deep connection between the first law of thermodynamics and the expansion law. But a proper generic connection between first law of thermodynamics and expansion law is absent in the literature.
 In 
%literature,
cosmology, 
%there are two kinds of 
%generalized 
the first law of thermodynamics is usually used either in the form 
%that are in use. The first one is 
$dE = TdS+WdV,$  the unified first law, or a form 
%and the other is one, 
without pressure term, as $-dE=TdS.$  The first form is applied to the whole volume within the horizon, in which 
%In the first one 
the energy is the Misner-Sharp energy, $E=\rho V$ contained within the given volume $V$ \cite{PhysRevD.75.084003}, while the second form is applied at the horizon, where energy $E$ refers to the flux crossing the apparent horizon of the Universe \cite{Cai_Kim}.  %In reference\cite{Mahith2018}, 
The expansion law was derived from both these forms of the first  law, in reference\cite{Mahith2018},  by explicitly using the Friedmann equations.
% as an additional input. 
%%%Our aim here is to obtain the expansion law from the thermodynamic law without the explicit use of Friedmann equation. On the other hand, we first obtain a common unique form by imposing both the forms of the thermodynamic laws at the horizon of an (n+1) dimensional universe, form which the expansion law for different gravity theories like Einstein, Gauss-Bonnet, Lovelock etc. follows naturally without the explicit assistance of the Friedmann equations.

Let us consider an (n+1) dimensional FRW Universe with metric 
%a null horizon as,
\begin{equation}\label{eq:metric}
ds^{2} = h_{ab}dx^{a}dx^{b} + a^{2}r^{2}d\Omega_{n-1}^{2},
\end{equation}
where $h_{ab}= diag\left[-1, a(t)^2/1-kr^2\right]$ is the two dimensional metric 
%projection to 
of the  $ t-r $ surface, $a$ is the scale factor of  expansion, $r$ is the co-moving radial distance, and $d\Omega_{n-1}$ is the metric of (n-1)-dimensional sphere with unit radius.  The spatial curvature constant have values 
$k = 1, 0 \, \textrm{and} \, -1,$ corresponding to a closed, flat and open Universe and, $a(t)$ is the scale of expansion. The apparent horizon of the Universe satisfies the condition,
%can be obtained using the relation 
$h_{ab} \partial_a\tilde{r} \partial_b\tilde{r}=0,$ (where $\tilde{r}= a(t)r$), which gives  the apparent horizon radius as,
\begin{equation}
\tilde{r}_A = \frac{1}{\sqrt{H^2 + \frac{k}{a^2}}},
\end{equation}
where $H$ is the Hubble parameter. From the standard relation for the surface gravity $\kappa$, the horizon temperature can have the form
% The temperature of the horizon is given by $T = \frac{\kappa}{2\pi},$ where $\kappa$ is the surface gravity of the horizon. From the standard relation for the surface gravity, the horizon temperature  can have the form
%which takes the form
\cite{Hayward_1998,Hayward:1998ee},
\begin{equation}\label{eq:apparent_radius}
T = \frac{1}{2\pi} \left[-\frac{1}{\tilde{r}_A} \left(1-\frac{\dot{\tilde{r}}_A}{2H\tilde{r}_A}\right)\right],
\end{equation}
%For the flat FRW Universe, the apparent horizon became the Hubble horizon with radius $\tilde{r}_{A} = 1/H$. The cosmic component is assumed to be a 
where the over-dot represents a derivative with respect to cosmic time.

Let us formulate the unified first law at the apparent horizon. 
The cosmic component is assumed to be a 
perfect fluid, such that the time and spacial components of the energy-momentum tensor are, $T^{0}_{0}=-\rho;~T^{i}_{i}= p$ with density $\rho$ and pressure $p$ of the cosmic
components. Thus the energy within the volume $V$ enclosed by the apparent horizon is $E=\rho V.$ Then the unified first 
%kind of the first 
law can be expressed as \cite{PhysRevD.75.084003},
%On this metric we have the first law $\tilde{T}dS= d\tilde{E} -WdV $ have the form \cite{}
%\begin{equation}
%\frac{-H}{2\pi}\left(1+ \frac{\dot{H}}{2H^{2}}\right)dS = \left(\rho dV + Vd\rho\right) -\frac{\left(\rho - p\right)}{2}dV.
%\end{equation}
\begin{equation}
 \frac{1}{2\pi} \left[-\frac{1}{\tilde{r}_A} \left(1-\frac{\dot{\tilde{r}}_A}{2H\tilde{r}_A}\right)\right]dS = \left(\rho dV + Vd\rho\right) -\frac{\left(\rho - p\right)}{2}dV.
\end{equation}
Using the continuity equation in (n+1) FRW Universe, $\dot{\rho} + nH\left(\rho +p \right)$ = 0, the above equation becomes
%\begin{equation}
%\frac{H}{2\pi}\left(1+ \frac{\dot{H}}{2H^{2}}\right)\dfrac{dS}{dt} = \frac{n\Omega_{n}}{H^{n-1}}\left(\rho + p\right)\left(1+ \frac{\dot{H}}{2H^{2}}\right)
%\end{equation} 
\begin{equation}
\label{eq:ds/dt}
-\dfrac{1}{2\pi H \tilde{r}_A^{n+1}}\frac{dS}{dt} =  -n\Omega_{n}\left(\rho + p\right).
\end{equation}
In obtaining the above relation, 
%here 
we took $V = \Omega_{n}\tilde{r}_A^{n}$, 
%as 
the 
%areal 
volume of (n+1) FRW Universe enclosed by the apparent horizon, where $\Omega_{n}$ is the areal volume of an n-dimensional sphere with unit radius. 

Now let us turn to the second form of the law, $-dE =TdS$ and see how it reduces at the horizon of an (n+1)-dimensional Friedmann Universe. It should be noted that, unlike in the previous form of the law, the energy $dE$ here is 
%by this law the observer is 
%referred to 
the energy flux across the apparent horizon. The observer measuring this flux is located on the apparent horizon, 
%and he is local to the apparent horizon, 
for whom 
% and not considering the the changes in the volume within the horizon. As a result the
the apparent horizon is virtually stationary. 
%As a result, 
Relative to this local observer, the temperature of the apparent horizon becomes $T=1/2\pi \tilde{r}_A $. 
%In FRW Universe, 
The 
%energy 
%supply vector for the perfect fluid have the form \cite{AKBAR20067},
%\begin{equation}
%\psi_{a} = \left(-\frac{1}{2}\left(\rho+p\right)H\tilde{r}, \frac{1}{2}\left(\rho+p\right)a\right),
%\end{equation}
%and the 
energy flux through the apparent horizon 
%of area, $A= n\Omega_{n}\tilde{r}_{A}^{n-1}$ 
during a small interval of time $dt$ is given by\cite{AKBAR20067} 
\begin{equation}
-dE= 
%A\psi_a= 
A\left(\rho+p\right)H\tilde{r}_A dt,
\end{equation}
where $A= n\Omega_{n}\tilde{r}_{A}^{n-1}$ is the area of the horizon of an (n+1) dimensional FRW Universe.
%The total energy crossing the horizon should be equivalent to $TdS$, the 
Then the first law 
%of thermodynamics of the form $-dE =TdS$ 
%in flat FRW universe with 
at the apparent horizon of the FRW Universe will take the form,
\begin{equation}
A\left(\rho+p\right)H\tilde{r}_A dt=\frac{1}{2\pi \tilde{r}_A}dS.
\end{equation}
%Once we put back the 
On substituting the area of the apparent horizon with some suitable rearrangements, the above equation becomes exactly similar to 
%one can see that the above equation cab be reduced to  
%horizon area with radius $r_{A}=1/H$  in to the above first law, the resulting expression easily reduces to 
Eq. (\ref{eq:ds/dt}), obtained for the unified form of the law of thermodynamics. The fact that Eq. (\ref{eq:ds/dt}) represents both forms of the thermodynamic laws at the horizon, is not surprising. However we use this result in our later considerations.
%Hence we showed that the two kinds of first law could be reduced to the same form
% follows the same relation
%(\ref{eq:ds/dt}) at the apparent horizon of an (n+1) dimensional FRW Universe.
% seems more fundamental.

%The important point is that 
%We have thus shown that the two kinds of the first law of thermodynamics 
%having different form of temperature and the change in energy follows to the 
%reduces to the same equation (\ref{eq:ds/dt}) at the Hubble horizon of the FRW universe. 
The integration of Eq. (\ref{eq:ds/dt}) using the continuity equation, $\dot\rho+nH(\rho+p)=0$  leads to
\begin{equation}\label{eq:integraleqn}
\frac{-1}{\pi}\int \frac{1}{\tilde{r}_A^{n+1}}dS = 2\Omega_{n}\left(\rho + \rho_{\Lambda}\right).
\end{equation} 
Here the dark energy density $\rho_{\Lambda}$ arises naturally as the integration constant and is equivalent 
%corresponding 
to the cosmological constant. It should be noted that, in reference 
%Which is the peculiar feature of the thermodynamic formulation of gravity 
\cite{2014GReGr}, Padmanabhan have obtained the cosmological constant density as an integration constant, following the principles of the emergent gravity paradigm. Combining the Eqs. (\ref{eq:ds/dt}) and (\ref{eq:integraleqn}), multiplay both the sides by $\dfrac{4\pi l_{P}^{n-1} }{n-2}H\tilde{r}_A^{n+2 }$ and rearranging suitably, will lead to
%of Eq.(\ref{eq:integraleqn})
\begin{widetext}	
\begin{equation}
\alpha \frac{4l_{P}^{n-1}\tilde{r}_A}{(n-1)}\dfrac{dS}{dt} = l_{P}^{n-1}\frac{\tilde{r}_A}{H^{-1}}\left[-\alpha\frac{2}{(n-1)} 4 \tilde{r}_A^{n+1}\int \frac{1}{\tilde{r}_A^{n+1}}dS + \left(\frac{\left[(n-2)\rho+np - 2\rho_{\Lambda}\right] V}{(n-2)}\right)\left(\frac{1}{4\pi\tilde{r}_A}\right)^{-1}\right].
\end{equation}
\end{widetext}
Where $\alpha = \dfrac{(n-1)}{2(n-2)}$. The first part of the second term on the right-hand side of the above equation is 
%arises due to the contribution from the bulk volume with the horizon of the universe and can be conveniently 
%Now we can identify the last term in the R.H.S. of the above equation 
%identified as the bulk $DoF$, $N_{bulk}$ corresponding to the equipartition of 
proportional to the Komar energy, while the second part is inversely proportional to the temperature as, $ \left(1/4\pi\tilde{r}_A\right)^{-1} = \left((1/2) k_B T\right)^{-1}$ 
%\\  at the horizon temperature, $k_{B}T=1/2\pi \tilde{r}_A$ 
\cite{PhysRevD.81.124040, cai, Sheykhi2013}. Hence the whole second term on the right-hand side can be conveniently identified as the DoF of the bulk within the apparent horizon, $N_{bulk}.$ 
% and by ansatz, 
Then the first term on the right-hand side can be the 
%the first term can be treated as the 
surface $DoF$, $ N_{sur}  $. 
%Then 
And the above equation can then be re-expressed as,    
\begin{equation}\label{eq:G.expansionlaw_equilibrium}
\alpha \frac{4l_{P}^{n-1}\tilde{r}_A}{(n-1)}\dfrac{dS}{dt} = l_{P}^{n-1}\frac{\tilde{r}_A}{H^{-1}}\left(N_{sur} - \epsilon N_{bulk}\right).
\end{equation}
Where $N_{sur}$ and $N_{bulk}$ are in the form,
\begin{align} \label{eq:gen surDoF}
N_{sur}&= -\alpha\frac{2}{(n-1)}4 \tilde{r}_A^{n+1}\int \frac{1}{\tilde{r}_A^{n+1}}dS  \quad\text{  and}\\\label{eq:gen bulkDoF} N_{bulk}&=-\epsilon\left(\frac{\left[(n-2)\rho+np - 2\rho_{\Lambda}\right] V}{(n-2)}\right)\left(\frac{1}{4\pi\tilde{r}_A}\right)^{-1}.
\end{align}
%and are the conventional degrees of freedom on the horizon surface and in bulk, respectively.
%and are compatible with their original definitions for Einstein's gravity.
%Using the entropy-area relation in different gravity theories, the above 
%Expansion law given in 
Here the equation for $N_{sur},$ is expressed as an integral over entropy. 
%a more general form for the surface DoF. 
For instance, in the (n+1) Einstein's gravity, $S= A/4l_{P}^{n-1}.$ On substituting this into the equation for $N_{sur}$ and using $A=n\Omega_n \tilde{r}_A^{n-1},$ it can easily be shown that $N_{sur}=  \alpha A/l_{P}^{n-1}$ is the relation for 
%the corresponding surface DoF. \\
the horizon surface DoF used previously\cite{Sheykhi2013}. 
%Equation (\ref{eq:G.expansionlaw_equilibrium}) with DoF in equations (\ref{eq:gen DoF}) is a more general form of the equation of emergence, and the corresponding degrees of freedom and are obtained using just the conservation equation and without the explicit use of the Friedmann equation. This 
%% and
%%applicable to any gravity theory. For a given gravity theory one have to use the corresponding  entropy-area relation. 
%%provided one must use the correct can be reduced to the expansion law corresponding to the respective gravitational theories in (n+1) dimensional flat FRW universe. 
%%We will now show that, 
%%the general law of expansion (\ref{eq:G.expansionlaw_equilibrium}) 
%%it 
%will 
%%suitably 
%reduces to the respective law of emergence for Einstein's gravity and other general gravity theories, already exists in literature.  
%the left hand side of equation (\ref{eq:G.expansionlaw_equilibrium}) is equivalent that formulated in reference \cite{Sheykhi2013,cai}, in terms of area. But the present form, has the additional advantage that it can be readily used even in non-equilibrium situation, with proper form for the effective entropy, which we discuss in the next section. 
%For Einstein's gravity in (n+1) dimension, 
%If we put 
%$S= A/4l_{P}^{n-1}$ 
%form which one can obtain $dS$, 
%then 
The L.H.S. of Eq. (\ref{eq:G.expansionlaw_equilibrium}) will reduce to $\alpha (dV/dt)$ for the same gravity. 
%and the $N_{sur}$ in equation (\ref{eq:gen 	DoF}) will be reduced to $N_{sur} = \alpha A/l_{P}^{n-1}$. 
As a result, the equation of emergence will be reduced to the standard form \cite{paddy2012jun,cai, Sheykhi2013}.
\begin{equation}\label{eq:expansionlaw}
\alpha \frac{dV}{dt} =  l_{P}^{n-1} \frac{\tilde{r}_A}{H^{-1}} \left(N_{sur} - \epsilon N_{bulk}\right).
\end{equation}
%provided one must use the correct can be reduced to the expansion law corresponding to the respective gravitational theories in (n+1) dimensional flat FRW Universe. 
%We will now show that, the general law of expansion (\ref{eq:G.expansionlaw_equilibrium}) will suitably reduces to the respective equation for Einstein's gravity and other general gravity theories, already exists in literature. 
%For Einstein's gravity in (n+1) dimension, 
%If we put 
%$S= A/4l_{P}^{n-1}$ form which one can obtain $dS$, then the L.H.S. of the above equation will reduces to the rate of increase in the Hubble volume as,
%\begin{equation}\label{eq:expansionlaw}
%\alpha \frac{dV}{dt} =  l_{P}^{n-1} \left(N_{sur} - \epsilon N_{bulk}\right),
%\end{equation}
%where the surface DoF became $N_{sur} = \alpha A/l_{P}^{n-1}$ and bulk DoF remains the same as in equation (\ref{eq: gen DoF}). The above equation is the same as the generalized expansion law proposed by Cai\cite{cai} for the Einstein gravity in flat (n+1) FRW Universe to derive the 
%% the expansion law also reduces to the (n+1) dimensional 
%Friedmann equation. This 
%%Which 
%shows the more general nature of the 
%%validity of the generalized 
%equation (\ref{eq:G.expansionlaw_equilibrium}), which we have deduced from the first law of thermodynamics. 
%
%Now we can check the consistency of the relation (\ref{eq:G.expansionlaw_equilibrium}) 
In the case of Gauss-Bonnet gravity, the entropy is of the form  \cite{Sheykhi2013,PhysRevD.65.084014, PhysRevD.69.104025},
%\begin{equation}\label{eq:entropy_gaussbonnet}
%S= \frac{A_{+}}{4l_{P}^{n-1}}\left(1 + \frac{n-1}{n-3}\frac{2\tilde{\alpha}}{r_{+}^{2}}\right).
%\end{equation}
\begin{equation}\label{eq:entropy_gaussbonnet}
S= \frac{A}{4l_{P}^{n-1}}\left(1 + \frac{n-1}{n-3}\frac{2\tilde{\alpha}}{\tilde{r}_{A}^{2}}\right).
\end{equation}
which having an additional correction term in comparison with that of Einstein's gravity.
The corresponding 
%For Gauss-Bonnet gravity 
surface DoF can then be obtained using Eq. (\ref{eq:gen surDoF}) as,
%have the form,
%\begin{eqnarray}
%N_{sur}&= -\alpha\frac{2}{(n-1)}\frac{4}{H^{n+1}}\int H^{n+1}dS \\
%  &= \frac{\alpha n \Omega_{n}}{H^{n+1}l_{P}^{n-1}}\int \left(1+ 2\tilde{\alpha}H^{2}\right)2HdH \\
%  & = \frac{\alpha n \Omega_{n}}{H^{n+1}l_{P}^{n-1}}\int \left(1+ 2\tilde{\alpha}H^{2}\right)2HdH\\
%  & = \frac{\alpha A_{+}}{l_{P}^{n-1}}\left[1+ \tilde{\alpha}H^{2}\right]
%\end{eqnarray} 
%\begin{equation}\label{eq:nsur_gaussbonnet}  
%N_{sur}= -\alpha\dfrac{2}{(n-1)}\dfrac{4}{H^{n+1}}\int H^{n+1}dS 
%= \dfrac{\alpha A_{+}}{l_{P}^{n-1}}\left[1+ \tilde{\alpha}H^{2}\right].
%\end{equation}
\begin{align}\label{eq:nsur_gaussbonnet}  
N_{sur} &= -\alpha\frac{2}{(n-1)}4 \tilde{r}_A^{n+1}\int \frac{1}{\tilde{r}_A^{n+1}}dS \nonumber\\
&= \dfrac{\alpha n\Omega_{n}\tilde{r}_{A}^{n-1}}{l_{P}^{n-1}}\left[1+ \tilde{\alpha}\tilde{r}_{A}^{-2}\right].
\end{align} %non flat is only obtained to this point need further modifications...................the above entropy S, $N_{sur}$ (\ref{eq:entropy_gaussbonnet}, \ref{eq:nsur_gaussbonnet}) respectively and
Now by using  the $N_{bulk}$ as given in the second equality of Eq. (\ref{eq:gen bulkDoF}), the generalized expansion law in Gauss-Bonnet gravity can be obtained as,  %(\ref{eq:G.expansionlaw_equilibrium}) has the form, 
%\begin{equation}
%\left(1+2\tilde{\alpha}H^{2}\right)\dot{H} + \left(1+\tilde{\alpha}H^{2}\right)H^{2}= - \frac{8\pi l_{P}^{n-1}}{n(n-1)}[(n-2)\rho + np].
%\end{equation}
\begin{align}\label{eq: expansionlaw_gaussbonnet}
 \left(1+\tilde{\alpha}\tilde{r}_{A}^{-2}\right)\tilde{r}_{A}^{-2} &-\left(1+2\tilde{\alpha}\tilde{r}_{A}^{-2}\right)\dot{\tilde{r}}_{A} H^{-1}\tilde{r}_{A}^{-3} =\nonumber\\ & - \frac{8\pi l_{P}^{n-1}}{n(n-1)}[(n-2)\rho + np-2\rho_{\Lambda}].
\end{align}
Which is same as the expansion law in Gauss-Bonnet gravity, obtained in \cite{Sheykhi2013}. 
%Multiplying both sides of the above equation by the factor $2a\dot{a}$. Then, integration of the above equation using the continuity equation $\dot{\rho} + nH\left(\rho +p\right) =0 $ will give the first Friedmann equation in Gauss-Bonnet gravity.  
Similarly, for 
more general 
Lovelock gravity the entropy has the form \cite{CAI2004237,Sheykhi2013},
\begin{equation}\label{eq:entropy_lovlock}
S = \dfrac{A}{4l_{P}^{n-1}}\sum_{i=1}^{m}\dfrac{i(n-1)}{(n-2i+1)}\hat{c}_{i}\tilde{r}_{A}^{2-2i}
\end{equation}
corresponding to which the surface $DoF$ (\ref{eq:gen surDoF}) 
%in Lovelock gravity has the 
takes the form,
\begin{equation}\label{eq:nsur_lovlock}
N_{sur} = \dfrac{\alpha A}{l_{P}^{n-1}}\sum_{i=1}^{m}\hat{c}_{i}\tilde{r}_{A}^{2-2i} .
\end{equation}
With the above forms of the surface $DoF,$
%( \ref{eq:nsur_lovlock}) 
%and entropy (\ref{eq:entropy_lovlock} 
the generalized expansion law (\ref{eq:G.expansionlaw_equilibrium}) in Lovelock gravity is 
\begin{align} \label{eq: expansionlaw_lovlock}
\sum_{i=1}^{m} \left(\hat{c}_{i}\tilde{r}_{A}^{-2i}\right. & \left. - i\hat{c}_{i}\tilde{r}_{A}^{-2i-1}\dot{\tilde{r}}_{A}H^{-1}  \right)  = \nonumber\\ &- \frac{8\pi l_{P}^{n-1}}{n(n-1)}\left[(n-2)\rho + np - 2\rho_{\Lambda}\right].
\end{align}
 From the above two equations of expansion law (\ref{eq: expansionlaw_gaussbonnet}) and (\ref{eq: expansionlaw_lovlock}), the respective Friedmann equations can be obtained, first by
 % The equation, 
 multiplying both sides of the 
 %above 
 equations with the factor $2a\dot{a}$ 
 %gives the first Friedmann equation in the Lovelock gravity on 
 and then integrate the result using the continuity equation \cite{Sheykhi2013}. 
% \textbf{Now we can define $A_{eff} = 4l_{P}^{n-1}S$, which  has the dimensions of the (n-1)area of the hyper surface in general and reduces to area of the apparent horizon in the (n+1) Einstein gravity. Then the surface degrees of freedom can be expressed as an integral over this effective area,
% \begin{equation}
% N_{sur}= -\alpha\frac{2}{(n-1)} \frac{\tilde{r}_A^{n+1}}{l_{p}^{n-1}}\int \frac{1}{\tilde{r}_A^{n+1}}dA_{eff}.
% \end{equation} 
%  Similarly, the R.H.S. of Eq. (\ref{eq:G.expansionlaw_equilibrium}) can be expressed as $\alpha \dfrac{\tilde{r}_A}{(n-1)}\dfrac{dA_{eff}}{dt}$. This is exactly same as the generalization of the expansion law introduced by Cai for Gauss-Bonnet and Lovelock gravity in flat FRW universe \cite{cai}, latter extended by Sheykhi in non-flat FRW universe \cite{Sheykhi2013}. Hence the generic derivation of the expansion law from the generalized first law gives a physical motivation to these generalizations.}
 
 The  
 expansion law we have obtained in Eq. (\ref{eq:G.expansionlaw_equilibrium}) (with Eqs. (\ref{eq:gen surDoF}, \ref{eq:gen bulkDoF})) have the following advantages. First, it can be taken as the general form of the expansion law, from which the expansion law for different gravity theories can be obtained by using the respective form of entropy. Second, it naturally selects what is known as areal volume instead of proper invariant volume, hence eliminating the discrepancy in the use of proper invariant volume. There exists a discrepancy in choosing the volume of the horizon in expressing the expansion law.
 % if it is expressed in a form involving the derivative of the volume. 
 %The previous generalization of the expansion law into the non-flat Universe \cite{Sheykhi2013} also results in the disparity, which is about the choice of volume in the expansion law. 
 The expansion law in a non-flat Universe can not be properly formulated using the proper invariant volume, but it can only be done using the areal volume. This is discussed in reference \cite{Hareesh_2019, Chang-Young:2013gwa}. In our derivation, since it is written directly in terms of entropy, Eq. (\ref{eq:G.expansionlaw_equilibrium}) naturally selects the areal volume rather than the proper invariant volume and avoid such disparity.
% \textbf{Similarly, the geometric generalization of $\dfrac{dV}{dt}$ proposed by Cai \cite{cai}, using the relation $\dfrac{dV}{dA}=\dfrac{R}{n-1}$  is criticized in reference \cite{FARAGALI}. Where the author questioning such generalizations from a mathematical point of view and suggest an alternative generalization. But the dynamical equations deduced from this alternative generalization shows deviation from the Friedmann equations in the Gauss-Bonnet gravity, where the author balanced the deviations using the generalized uncertainty principle \cite{FARAGALI}. The generic derivation of the expansion law from the generalized first law gives a physical motivation to the generalization proposed by Cai.} 
 Thirdly, the more prominent  advantage is that this form of the expansion law can be easily generalized to the non-equilibrium thermodynamic situations, which we consider in the next section.
 In the next section, we use the same idea to derive the expansion law in the general
 gravity models like $ f(R) $ gravity, including non-equilibrium description.  

 It should also be noted that, in formulating a general law of expansion in flat FRW universe, Cai has postulated the relation for the rate of change of the Hubble volume in both Gauss-Bonnet and Lovelock gravities as $\alpha \dfrac{1}{(n-1)H}\dfrac{dA_{eff}}{dt}$, in order to arrive at the Friedmann equation in both these gravity theories \cite{cai}. Here  the effective area, $A_{eff} = 4l_{p}^{n-1}S$. Latter Sheykhi extended the law of expansion to non-flat FRW Universe by postulating  $\alpha \dfrac{\tilde{r}_A}{(n-1)}\dfrac{dA_{eff}}{dt}$ \cite{,Sheykhi2013}. In our approach, the L.H.S of equation (\ref{eq:G.expansionlaw_equilibrium}), which is a consequence from the first law of thermodynamics, will naturally reduce to a form similar to the one postulated by Cai and Sheykhi, hence giving it a basic physical motivation.
\section{Expansion law from first law of thermodynamics - non equilibrium case}\label{sec.3}

 In this section, we extend the above procedure to obtain the expansion law from the generalized first law in a non-equilibrium situation. It has been shown that thermodynamics of gravity theories with higher-order curvature corrections requires a non-equilibrium treatment. The field equations in such theories can be derived using a modified entropy balance relation $dS = \delta Q/T + d_i S$ \cite{PhysRevLett.96.121301}, where $d_i S$ is the entropy generated due to the system being out of equilibrium. The $f(R)$
gravity is the typical example for theory with higher-order curvature correction, in which action is an arbitrary function of curvature scalar, $R$ \cite{Nojiri-Odintsov:2011, Nojiri-Odintsov-vko:2017, Sotiriou:2008rp}. 
%In expressing the Friedmann equations in $f(R)$ gravity as generalised first law of thermodynamics at the horizon, 
Akbar and Cai \cite{AKBAR2007243} have found that 
%on expressing 
the Friedmann equations of $f(R)$ gravity on the horizon assumes the 
%as first law of thermodynamics, one indeed arrive at the 
non-equilibrium first law of thermodynamics,
%the first law indeed takes the non-equiibrium form, 
\begin{equation} \label{eq:non-equilibrium firstlaw}
dE = TdS+WdV+Td_iS.
\end{equation}
This indicates that it is needed to use 
% Hence we have to use the generalized first law in the 
the non-equilibrium version of the generalized first law to derive the expansion law in gravity theories like $f(R).$ 
%  The generalized first law in the equilibrium description is not valid in the gravity models with higher order curvature correction, this is due to the entropy production arises from the non equilibrium nature of such models \cite{PhysRevLett.96.121301}.Hence we have to use the generalized first law in the non-equilibrium description to derive the general expansion law for the gravity models with higher order correction terms. 

For a general derivation of the expansion law in non-equilibrium, we choose the unified formulation for the modified gravity \cite{PhysRevD.90.104042}, which accounts for the non-equilibrium characteristics of the modified gravities by adopting the effective coupling strength $ G_{eff} $ instead of $G$ in the Einstein gravity.
%On this formalism the extra terms in the field equation are sum up to the stress energy tensor, $T_{\mu\nu}^{(eff)} = T_{\mu\nu}^{(m)} + T_{\mu\nu}^{(MG)} $, and hence the effective total density is  $\rho_{t} = \rho_{m} + \rho_{MG}$ and pressure $p_{t} = p_{m} + p_{MG}$. Then the field equation have the form,
On this approach, the field equation in the gravity theories having higher curvature corrections can generally be expressed as
  \begin{equation}\label{eqn:einfr}
  G_{\mu\nu} \equiv R_{\mu\nu}-\frac{1}{2}Rg_{\mu\nu} = 8\pi G_{eff}T_{\mu\nu}^{(eff)}.
  \end{equation} 
where $G_{eff}$ depends on time in general and $T_{\mu\nu}^{(eff)}$ is an effective stress-energy tensor, which can be bifurcated as, $T_{\mu\nu}^{(eff)} = T_{\mu\nu}^{(m)} + T_{\mu\nu}^{(MG)}. $ Here 
%with 
$T_{\mu\nu}^{(m)}$ 
%the tensorial part 
is due to matter and $T_{\mu\nu}^{(MG)}$ is that arises due to the modified gravity part, which reflects 
% due to 
the dependence on the higher-order curvature. The density and pressure 
%corresponding density and pressure  
can also be bifurcated into the matter and modified gravity part as, 
%expressed as, 
$\rho_{t} = \rho_{m} + \rho_{MG}$ and pressure $p_{t} = p_{m} + p_{MG}$.

%In such a case, time component of generalized energy momentum conservation $\nabla_{\mu}G^{\mu}_{\,\nu} = 8\pi\nabla_{\mu}\left(G_{fee}T^{\mu\,\left(eff\right)}_{\nu}\right)$ should give 
The generalized continuity equation in these kinds of gravity theories can be suitably be expressed as \cite{PhysRevD.90.104042}, 
  \begin{equation}\label{eq: Gcontiniutyeqn_non-equilibrium}
  \dot{\rho_{t}} + nH\left(\rho_{t} +p_{t}\right)=-\dfrac{\dot{G}_{eff}}{G_{eff}}\rho_{t}.
  \end{equation}
The 
%non-zero 
term in the R.H.S of Eq. (\ref{eq: Gcontiniutyeqn_non-equilibrium}) is the energy dissipation arises due to the non-equilibrium, which balances the energy flow. The above continuity equation can effectively be rewritten as,
  \begin{equation}\label{eq:modified_continiuityeqn}
\dot{\rho_{t}} + nH\left(\rho_{t} +p_{eff}\right)= 0,
\end{equation}
where $p_{eff} \equiv p_{t}+\dfrac{\dot{G}_{eff}}{nHG_{eff}}\rho_{t}$ can be considered as the effective pressure in the non equilibrium description having the energy dissipation. In literature, a similar idea was used in the bulk viscus model of the Universe to define the so-called effective pressure \cite{Weinberg}.

 Now lets assume the (n+1) FRW Universe having the metric as in Eq. (\ref{eq:metric}), the non-equilibrium first law (\ref{eq:non-equilibrium firstlaw}) takes the form,
\begin{equation}
\rho_{t} dV + Vd\rho_{t} = \dfrac{-1}{2\pi \tilde{r}_A } \left(1-\dfrac{\dot{\tilde{r}}_A}{2H\tilde{r}_A}\right)d\bar{S}+\dfrac{\left(\rho_{t} - p_{eff}\right)}{2}dV,
\end{equation}
where we took the total change in energy inside the horizon as $E=\rho_{t} V,$  the temperature of the horizon as, $T=\frac{-1}{2\pi \tilde{r}_A } \left(1-\dfrac{\dot{\tilde{r}}_A}{2H\tilde{r}_A}\right), $ effective work density  $W=\dfrac{\left(\rho_{t} - p_{eff}\right)}{2}$ and $d\bar{S} = dS +d_iS$.  Using the continuity equation (\ref{eq:modified_continiuityeqn}), the above equation can be reduced to the simple form, 
\begin{equation} \label{eq:d(bar{s})/dt}
-\dfrac{1}{2\pi H \tilde{r}_A^{n+1}}\frac{d\bar{S}}{dt} =  -n\Omega_{n}\left(\rho_{t} + p_{eff}\right),
\end{equation}
where used $V= \Omega_{n}\tilde{r}_{A}^{n}$ as in the last section.

Next we consider 
%Similarly, The 
first law of the form 
\begin{equation}\label{TdS12}
-dE =T \left(dS + d_iS\right) = Td\bar{S}
\end{equation}
%$-dE =T \left(dS + d_iS\right) = Td\bar{S}$ 
in FRW Universe with $dE$ as the energy flux $A\left(\rho_{t}+ p_{eff}\right)H\tilde{r}_Adt$ for the horizon with temperature $T=1 /2\pi\tilde{r}_A.$ With little algebra it can show that, this form of the first will leads to the same as in Eq. (\ref{eq:d(bar{s})/dt}) at the horizon.
 %Similarly, The first law of the form $-dE =Td\bar{S}$ in (n+1) FRW Universe with $dE$ as the energy flux $A\left(\rho_{t}+ p_{eff}\right)H\tilde{r}_Adt$ for the horizon temperature $T= /2\pi\tilde{r}_A$ leads to the above equation (\ref{eq:d(bar{s})/dt}), shows the consistency of our non-equilibrium formulation. The above equation (\ref{eq:d(bar{s})/dt}) is very similar to the equation (\ref{eq:ds/dt}) in the equilibrium description, where the change in entropy $dS$, density $\rho$ and pressure $p$ are replaced with the total change in entropy $d\bar{S}$ including the entropy production $d_{i}S$, total density $\rho_{t}$ and effective pressure in the non-equilibrium description $p_{eff}$ respectively.
 
  Using the continuity equation (\ref{eq:modified_continiuityeqn}), the integral form of the above equation can be expressed as
\begin{equation}\label{eq:integraleqn2}
\frac{-1}{\pi}\int \frac{1}{\tilde{r}_A^{n+1}}d\bar{S} = 2\Omega_{n}(\rho_{t} +\rho_{\Lambda}).
\end{equation} 
Combining the Eqs. (\ref{eq:d(bar{s})/dt}) and (\ref{eq:integraleqn2}) and multiplay both sides by $\dfrac{4\pi l_{P}^{n-1} }{n-2}H\tilde{r}_A^{n+2 }$ will leads to
\begin{equation}\label{eq:G.expansionlaw non-equilibrium}
\alpha \frac{4l_{P}^{n-1}\tilde{r}_A}{(n-1)}\dfrac{d\bar{S}}{dt} = l_{P}^{n-1}\frac{\tilde{r}_A}{H^{-1}}\left(N_{sur} - \epsilon N_{bulk}\right),
\end{equation}
with the DoF on surface, $N_{sur}$ and that of bulk, $N_{bulk}$ as
\begin{align} \label{eq: gen surDoF_non-equilibrium}
N_{sur}&=-\alpha\frac{2}{(n-1)}4 \tilde{r}_A^{n+1}\int \frac{1}{\tilde{r}_A^{n+1}}d\bar{S}  \quad\text{and }\\ \label{eq: gen bulkDoF_non-equilibrium} N_{bulk}&=-\epsilon\left(\frac{\left[(n-2)\rho_{t}+np_{eff} - 2\rho_{\Lambda}\right] V}{(n-2)}\right)\left(\frac{1}{4\pi\tilde{r}_{A}}\right)^{-1}. 
\end{align}
The Eq. (\ref{eq:G.expansionlaw non-equilibrium}) 
%(\ref{eq: gen DoF_non-equilibrium}) 
represents 
the expansion law in the non-equilibrium situation with  $DoF$ defined in Eqs. (\ref{eq: gen surDoF_non-equilibrium}, \ref{eq: gen bulkDoF_non-equilibrium}). 
%given in the non equilibrium description of (n+1) FRW Universe respectively. 
%From comparison, 
Interestingly, the above equation of the expansion law is similar in form
to the corresponding Eq. (\ref{eq:G.expansionlaw_equilibrium}) obtained for the equilibrium description. 
Even though the formal appearance looks the same, the entropy, density and pressure in the present equation are different in such a way that they contain the  
contribution due to the non-equilibrium nature. 
%In the next section, we apply the above  
%expansion law to the special case
%of $f(R)$ gravity in  (n+1) FRW Universe.
%
%The equation (\ref{eq:G.expansionlaw non-equilibrium}, \ref{eq: gen DoF_non-equilibrium}) represents the structure of more general form of the expansion law and the $DoF$ in the non equilibrium description of (n+1) FRW Universe respectively. The expansion law in different gravity models with higher order curvature correction terms can be obtained from the relation (\ref{eq:G.expansionlaw non-equilibrium}) using the area entropy relation. Since we used the unified formulation for the modified gravity using the effective coupling strength $G_{eff}$ the above derivation can be used to a set of gravity models  having the non equilibrium description. Hence we derived a general form of expansion law from the generalized fist law of thermodynamics in the non-equilibrium description. Here we only used the continuity equation in addition to the first law. In the next section we deduce the expansion law to the f(R) gravity in  (n+1) FRW Universe from the general structure of the expansion law (\ref{eq:G.expansionlaw non-equilibrium}), that we derived here.   

\section{Expansion law for f(R) gravity in (n+1) FRW Universe} \label{sec.4}

In this section, we derive the expansion law in the f(R) gravity using the generalized expansion law in the non-equilibrium description, obtained in the previous section. The f(R) gravity is one of the main candidates from the gravity models with higher-order curvature corrections \cite{Nojiri-Odintsov:2011, Nojiri-Odintsov-vko:2017, Sotiriou:2008rp}, which can explain the late acceleration of the Universe.
% However the f(R) gravity has the Wald entropy of the form $S= Af_{R}(R)/4G$, which is the non-equilibrium situation, hence we need to include the entropy production term $Td_{i}S$ in the generalized first law for f(R) gravity \cite{AKBAR2007243, PhysRevLett.96.121301},
%\begin{equation}
%dE =TdS+Td_{i}S + WdV.
%\end{equation}
%The entropy production terms are different in different formalism of the the first law. We can find mainly two kind of first laws for f(R) gravity in literature based on the different generalizations on the Misner-sharp energy $E$ that used to formulate the first law \cite{AKBAR2007243,unifiedformulationTian:2014ila}.
The Einstein-Hilbert action of f(R) gravity has the form, 
\begin{equation}
A = \int d^{(n+1)}x\, \sqrt{-g}\left(f(R) + 2\kappa L_{m}\right),
\end{equation}
where $\kappa = 8\pi G $. From the variational principle, $\delta A = 0 $ 
%yields 
the field equations can be obtained as given in Eq. (\ref{eqn:einfr}). The matter and gravity components of the effective energy-momentum tensor, $T_{\mu\nu}^{(eff)}$ are,
% \\ can be obtained as,
%\begin{equation}
%R_{\mu\nu}f^{\prime}(R) - \frac{1}{2}g_{\mu\nu}f(R)+ g_{\mu\nu}\nabla^{2}f^{\prime}(R)- \nabla_{\mu}\nabla_{\nu}f^{\prime}(R)  = 8\pi G T_{\mu\nu}^{(m)} 
%\end{equation}
%where $f^{\prime}(R)=d f(R)/dR$. In the unified formulation the field equations can be expressed as \cite{PhysRevD.90.104042},
%\begin{equation}
%G_{\mu\nu} = 8\pi G_{eff}\left[T_{\mu\nu}^{(m)} + T_{\mu\nu}^{(cur)}\right]\equiv 8\pi G_{eff}T_{\mu\nu}^{(eff)},
%\end{equation}
%where $G_{\mu\nu} = R_{\mu\nu}-\frac{1}{2}g_{\mu\nu}R$ is the Einstein tensor,  $G_{eff} = G/f^{\prime}(R)$ can be treated as the effective coupling strength in the f(R) gravity. The stress-energy tensor contributions from matter with density $\rho^{(m)}$ and pressure $p^{(m)}$ is defined as, 
\begin{equation}
T_{\mu\nu}^{(m)} = \left(\rho_m + p_m\right)U_{\mu} U_{\nu} + p_m g_{\mu\nu}
\end{equation}
and 
%the additional energy momentum tensor arises from higher order curvature terms is %\cite{PhysRevD.90.104042}
\begin{align}
&T_{\mu\nu}^{(MG)} = \frac{1}{8\pi G}\times \nonumber\\&\left(\frac{f(R)-Rf^{\prime}(R)}{2}g_{\mu\nu}+\nabla_{\mu}\nabla_{\nu}f^{\prime}(R) -  g_{\mu\nu}\nabla^{2}f^{\prime}(R) \right)
\end{align}
Where  $\rho_{m}$ is the 
%matter 
density of matter and $p_{m}$ is its pressure, 
%Therefore we can express the field equation in f(R) gravity as,
%\begin{equation}
%G_{\mu\nu} = 8\pi G_{eff}\left[T_{\mu\nu}^{(m)} + T_{\mu\nu}^{(cur)}\right] \equiv 8\pi G_{eff}T_{\mu\nu}^{(eff)}.
%\end{equation}
Then the total effective energy-momentum tensor implies a total density $\rho_{t}$ and total pressure $p_{t}$ as
%  $T_{\,0}^{0\, (eff)} = -\rho_{t}$ and $T_{\,i}^{i\, (eff)} = p_{t}$.
%Where the effective density $\rho_{t} $ and pressure $p_{t}$ as,
%Where the effective density $\rho_{t} = \rho_{m} + \rho_{MG}$ and pressure $p_{t} = p_{m} + p_{MG}$ as,
\begin{align} \label{eq:rho_t }
\rho_{t} = &\rho_{m} + \dfrac{1}{8\pi G}\left[\dfrac{ Rf^{\prime}(R)-f(R)}{2}- nH\dot{f}'(R)\right] \\ \label{eq: p_t}
p_{t} = &p_{m} + \dfrac{1}{8\pi G}\left[\dfrac{f(R) - Rf^{\prime}(R)}{2} + \ddot{f}'(R)\right. \nonumber\\ & ~~~~~~~~~~~~~~~~~~~~\quad \qquad + (n-1)H\dot{f}'(R)\bigg].
\end{align}
%The corresponding Friedmann equations in (n+1) FRW Universe are
%\begin{eqnarray}\label{eq: 1fieldeqn_f(R)}
%H^{2}+\frac{k}{a^2}  = &\dfrac{16\pi G_{eff}}{n(n-1)}\rho_{t} \\\label{eq: 2fieldeqn_f(R)}
%\dot{H} - \frac{k}{a^2} = &-\dfrac{8\pi G_{eff}}{(n-1)}\left(\rho_{t} + p_{t}\right).
%\end{eqnarray}
% The total energy $E = \rho_{t}V$ is equivalent to the Misner-Sharp energy  at the Hubble horizon of the form \cite{PhysRevD.90.104042},
%\begin{equation}
%E = \frac{f^{\prime}(R)}{2G H}\equiv \frac{1}{2G_{eff} H}.
%\end{equation}
%The above form of the generalized Misner-Sharp energy is seems to be geometrical sourced by the total effective energy momentum tensor,$T_{\mu\nu}^{eff}$ other than the total energy of the cosmological components, $\rho_{m}V$. 
 %Where the 
 The total effective density $\rho_{t}$ (\ref{eq:rho_t }) coupled to $G_{\mu\nu}$ through the 
 %act as the source density with an effective 
 coupling strength $G_{eff} = G/f^{\prime}(R)$. 
%Where 
The time evolution of the total density can be obtained as, 
%is in the form, 
\begin{equation}
\dot{\rho_{t}} = \dot{\rho_{m}} +\dfrac{1}{8\pi G}\left[\dfrac{ R\dot{f}'(R)}{2} -n\dot{H}\dot{f}'(R) -nH\ddot{f}'(R)   \right].
\end{equation}
%and
%\begin{equation}
%\rho_{t} +p_{t} = \rho_{m} + p_{m} + \dfrac{1}{8\pi G}\left[ -nH\dot{f}'(R)+\ddot{f}'(R)-H\dot{f}'(R) \right].
%\end{equation}
%Hence the continuity equation have the form,
%\begin{equation}
%\dot{\rho_{t}} + nH \left(\rho_{t} +p_{t} \right) = \dfrac{\dot{f}'(R)}{8\pi G}\left[\dfrac{R}{2} - n\dot{H} -nH \right]
%\end{equation}
%where we have taken for granted the ordinary cosmic components as the perfect fluid, $ \dot{\rho_{m}} + nH \left(\rho_{m} +p_{m} \right) = 0$. Now we put Ricci scalar in (n+1) FRW Universe, $R= 2n\left[\dot{H} + \dfrac{\left(n+1\right)}{2}H^{2}\right] $ and then we get the
%Using this 
%%and 
%the corresponding continuity equation can be obtained from the general expression (\ref{eq: Gcontiniutyeqn_non-equilibrium}) using the first field equation (\ref{eqn:einfr}) as,
% One can show that the above continuity equation in $f(R)$ gravity can be reduced to the general non -equilibrium continuity equation (\ref{eq: Gcontiniutyeqn_nonequilibrium})
Using this, one can obtain a continuity equation analogous to the general expression (\ref{eq: Gcontiniutyeqn_non-equilibrium}),    
\begin{equation}\label{eq:fRcontiniutyeqn_non-equilibrium}
\dot{\rho_{t}} + nH \left(\rho_{t} +p_{t} \right) = \dfrac{ n(n-1)\dot{f}'(R)}{16\pi G\tilde{r}_{A}^{2}}.
\end{equation}
The continuity equation shows dissipation characteristics of the effective total density $\rho_{t}$. The above continuity equation can be rewritten as
\begin{equation}
\dot{\rho_{t}} + nH \left(\rho_{t} +p_{eff} \right) = 0,
\end{equation}
where $p_{eff} \equiv p_{t} -\dfrac{ (n-1)\dot{f}'(R)}{16\pi GH\tilde{r}_{A}^{2}}$ is the effective pressure due to the energy dissipation. 
% The non equilibrium nature of f(R) gravity is very evident from the Horizon entropy in f(R) gravity. For the entropy of the black hole horizon in $f(R)$ gravity,  %$S$ of the Horizon in $f(R)$  gravity,
%% can be thought as the Noether's charge for the Kodama observer, 
%$S=Af^{\prime}(R)/4l_{p}^{n-1}$\cite{PhysRevD.48.R3427}, which is in the non-equilibrium situation in general due to the $f^{\prime}(R)$ term in the entropy  \cite{PhysRevLett.96.121301}.))))))))))))))000000000)))))

Now we will derive the expansion law for $f(R)$ gravity using the non-equilibrium first law (\ref{TdS12}). The total entropy change has two components; the first corresponds to change in the 
 Wald entropy of the horizon, 
 %in $f(R)$  gravity,
% can be thought as the Noether's charge for the Kodama observer, 
$S=Af^{\prime}(R)/4l_{p}^{n-1}$ \cite{PhysRevD.48.R3427} and the second is $d_iS,$ the additional entropy change generated due to non-equilibrium evolution.
% in which 
%itself reflects 
%the non-equilibrium situation is incorporated within $f^{\prime}(R)$\cite{PhysRevLett.96.121301}. 
%However starting with the 
%%The dynamic equation from the generalized 
%first law of thermodynamics 
%%$dE =TdS + WdV$ (or 
%$-dE = TdS$
%%) 
%on the apparent horizon 
%one will not reach in to the 
%%in $ f(R) $ gravity with the Wald entropy, the dynamic equation is inconsistent with the 
%respective Friedmann equation in $ f(R) $ gravity \cite{AKBAR20067}. On the other hand 
%Hence 
We need to include additional entropy production term $Td_{i}S$ in the generalized first law in the volume inside the apparent horizon for f(R) gravity  \cite{AKBAR2007243, PhysRevLett.96.121301},
\begin{equation}
%dE =TdS+Td_{i}S + WdV.
-dE=TdS+Td_iS.
\end{equation}
%The entropy production terms are different in different formalism of the the first law. We can find mainly two kind of first laws for f(R) gravity in literature based on the different generalizations on the Misner-sharp energy $E$ that used to formulate the first law \cite{AKBAR2007243,unifiedformulationTian:2014ila}
Where 
%$E =\rho_{t}V$ is the total effective Misner-Sharp energy inside the apparent horizon, $W=\dfrac{\rho-p_{eff}}{2}$ is the work density 
$-dE=A\left(\rho_t+p_{eff}\right)H\tilde{r}_A dt$ is flux across the horizon and $T=\dfrac{1}{2\pi \tilde{r}_A }$
%\left(1-\dfrac{\dot{\tilde{r}}_A}{2H\tilde{r}_A}\right)$ 
is the horizon temperature. 
%For an infinitesimal increment in the energy, $E$ causes to slight increment in the apparent horizon leading to the change in the total entropy, $d\bar{S}= dS+ d_{i}S$, Where  
The variation of the Wald entropy $S$ is, $dS= \frac{A}{4\l_{P}^{n-1}}\left[df^{\prime}(R) + \frac{(n-1)f^{\prime}(R)d\tilde{r}_{A}}{\tilde{r}_{A}}\right]$. Then the additional entropy generated, $d_iS$ can be obtained by substituting $dE$ and $dS$ in to the above form of the first law. A little algebra with the help of respective Einstein equations, it can be shown that.
%$Td_iS=-dE-TdS.$\\
%The additional entropy $d_{t}S$ arises due to the dynamic nature of the effective coupling strength, $G_{eff}=\dfrac{G}{f^{\prime}(R)}$. Which can be identified as 
%In addition to that, an entropy production $d_{i}S$ happence due to the non-equilibrium nature of the f(R) gravity, the first law have the form,
%\begin{equation}
%-dE = Td\bar{S} = TdS +Td_{i}S,
%\end{equation}
%where the energy flux  through the apparent horizon $-dE = A(\rho_{t}+p_{eff})dt$, the temperature, $T=1/2\pi\tilde{r}_{A}$ and the corresponding entropy change $dS= \frac{A}{4\l_{P}^{n-1}}\left[df^{\prime}(R) + \frac{(n-1)f^{\prime}(R)d\tilde{r}_{A}}{\tilde{r}_{A}}\right]$, hence the entropy production $d_{i}S$ can be estimated as
%\begin{equation}
%\frac{H}{2\pi}d_{i}S= A(\rho_{t}+p_{eff})dt - \frac{H}{2\pi}\frac{A}{4\l_{P}^{n-1}}\left[df^{\prime}(R) + \frac{(n-1)f^{\prime}(R)d\tilde{r}_{A}}{\tilde{r}_{A}}\right] ,
%\end{equation} 
%\begin{equation}
%\frac{H}{2\pi}d_{i}S= A(\rho_{t}+p_{t})dt -\dfrac{ (n-1)H\dot{f}'(R)}{16\pi G}Adt - \frac{H}{2\pi}\frac{A}{4\l_{P}^{n-1}}\left[df^{\prime}(R) + \frac{(n-1)f^{\prime}(R)d\tilde{r}_{A}}{\tilde{r}_{A}}\right] ,
%\end{equation}
%using the field equation (\ref{eq: 2fieldeqn_f(R)})where $d_{i}S$ can be identified as, 
\begin{equation}
d_{i}S =-\dfrac{(n+1)A}{8l_{P}^{n-1}}df^{\prime}(R).
\end{equation}
%where we have substitutes $\rho_t+p_{eff}$ using the Friedmann equation of the $f(R)$ gravity.
 %using the field equation(\ref{eq: 2fieldeqn_f(R)}). 
 Hence the total entropy change should be
\begin{equation}\label{eq: bar{S}}
d\bar{S} =\dfrac{(n-1)A}{4l_{P}^{n-1}}\left[\frac{f^{\prime}(R)d\tilde{r}_{A}}{\tilde{r}_{A}}-\frac{df^{\prime}(R)}{2}\right].
\end{equation}
Now we can express the expansion law in $f(R)$ gravity from the general expression (\ref{eq:G.expansionlaw non-equilibrium}) using Eq. (\ref{eq: bar{S}}) as 
\begin{equation}\label{eq:f(R)_expansionlaw non-equilibrium}
 \alpha A \tilde{r}_{A} \left[\frac{f^{\prime}(R)\dot{\tilde{r}}_{A}}{\tilde{r}_{A}}-\frac{\dot{f}'(R)}{2}\right] = l_{P}^{n-1}\frac{\tilde{r}_A}{H^{-1}}\left(N_{sur} - \epsilon N_{bulk}\right).
\end{equation}
Here we can estimate the $DoF$ associated with the horizon surface of the FRW Universe in $f(R)$ gravity using the general expression (\ref{eq: gen surDoF_non-equilibrium}),
\begin{equation}
N_{sur}
%&=& -\alpha\frac{2}{(n-1)}4 \tilde{r}_A^{n+1}\int \frac{1}{\tilde{r}_A^{n+1}}d\bar{S}\\
%&=& -\alpha\frac{2}{(n-1)}4 \tilde{r}_A^{n+1}\int \frac{1}{\tilde{r}_A^{n+1}}\dfrac{(n-1)A}{4l_{P}^{n-1}}\left[\frac{f^{\prime}(R)d\tilde{r}_{A}}{\tilde{r}_{A}}-\frac{df^{\prime}(R)}{2}\right]\\
= \alpha\frac{n\Omega_{n}}{l_{P}^{n-1}} \tilde{r}_A^{n+1}\int \left[-\frac{2f^{\prime}(R)d\tilde{r}_{A}}{\tilde{r}_{A}^{3}}+\frac{df^{\prime}(R)}{\tilde{r}_{A}^2}\right].
\end{equation}
%where 
The integrand in the above equation will reduces to the form $ d\left(\dfrac{f^{\prime}(R)}{\tilde{r}_{A}^2}\right)$, and hence the surface $DoF$ in $ f(R) $ gravity reduces to
\begin{equation}\label{eq:Nsur_f(R)}
 N_{sur}= \alpha\dfrac{Af^{\prime}(R)}{l_{P}^{n-1}}.
\end{equation}
For Einstein gravity, $f^{\prime}(R) = 1$ and hence have one $DoF$ in the unit Plank area $l_{p}^{n-1}$ seems a special case. But in general there have $f^{\prime}(R)$ number of $DoF$ associated with the unit Plank area  for $f(R)$ gravity in the (n+1) FRW Universe.  
%&=& \alpha\dfrac{Af^{\prime}(R)}{l_{P}^{n-1}}\label{eq:Nsur_f(R)},
Similarly, the number of bulk $DoF$ in $f(R) $ gravity can be obtained as
\begin{equation}\label{eq:Nbulk_f(R)}
N_{bulk}=-\epsilon\left(\frac{\left[(n-2)\rho_{t}+np_{eff}- 2\rho_{\Lambda}\right] V}{(n-2)}\right)\left(\frac{1}{4\pi\tilde{r}_{A}}\right)^{-1}.
\end{equation}
Then We can write the expansion law in f(R) gravity (\ref{eq:f(R)_expansionlaw non-equilibrium})  using the surface degrees of freedom, $N_{sur}$ in Eq. (\ref{eq:Nsur_f(R)}) and the bulk degrees of freedom, $N_{bulk}$ in f(R) gravity (\ref{eq:Nbulk_f(R)}) as,
%in the expansion law  in the non-equilibrium description \ref{eq:G.expansionlaw non-equilibrium} will be in the form,
\begin{widetext}
\begin{equation}\label{eq:f(R).expansionlaw non-equilibrium}
 \alpha A \tilde{r}_{A} \left[\frac{f^{\prime}(R)\dot{\tilde{r}}_{A}}{\tilde{r}_{A}}-\frac{\dot{f}'(R)}{2}\right] =l_{p}^{n-1}\tilde{r}_{A}H\left[\alpha Af^{\prime}(R) + \left(\frac{\left[(n-2)\rho_{t}+np_{eff}-2\rho_{\Lambda}\right] V l_{p}^{n-1}}{(n-2)}\right)\left(\frac{1}{4\pi\tilde{r}_{A}}\right)^{-1}\right].
\end{equation}
Which can be reduced to the form,
% to multiplying with $\frac{H^{2}}{\alpha A}$, then we can rearrange to the form,
\begin{equation}
\frac{\dot{\tilde{r}}_{A}}{H\tilde{r}_{A}^{3}}-\dfrac{1}{\tilde{r}_{A}^{2}}= \frac{\dot{f}'(R)}{2f^{\prime}(R)H\tilde{r}_{A}^{2}} + \frac{8\pi G_{eff}}{(n-1)} \left(\rho_{t}+p_{eff}\right) -\frac{16\pi G_{eff}}{n(n-1)} \left[\rho_{t}+\rho_{\Lambda}\right].
\end{equation}
\end{widetext}
%\begin{equation}
%\dot{H}f^{\prime}(R) +H^{2}f^{\prime}(R) = -\frac{8\pi l_{p}^{n-1}}{n(n-1)}\left[(n-2)\rho_{t} + np_{eff}-2\rho_{\Lambda} \right] -\frac{\dot{f}'(R)H}{2}.
%\end{equation}
On
%ce we put 
substituting back the effective pressure $p_{eff}= p_{t} -\dfrac{ (n-1)\dot{f}'(R)}{16\pi GH\tilde{r}_{A}^{2}}$ to the above expansion law, the first term in the R.H.S. of the above equation will cancelled out with the additional pressure term from the $p_{eff}$ and hence we finally get the expansion law in f(R) gravity as 
%\begin{equation}\label{eq:dynamicequation_f(R)}
%\dot{H} +H^{2} = -\frac{8\pi l_{p}^{n-1}}{n(n-1)f^{\prime}(R)}\left[(n-2)\rho_{t} + np_{t} -2\rho_{\Lambda}  \right].
%\end{equation} 
\begin{equation}\label{eq:dynamiceq_fR_rA}
-\frac{\dot{\tilde{r}}_{A}}{H\tilde{r}_{A}^{3}}+\dfrac{1}{\tilde{r}_{A}^{2}} =  \frac{16\pi G_{eff}}{n(n-1)} \left(\rho_{t}+\rho_{\Lambda}\right)- \frac{8\pi G_{eff}}{(n-1)} \left(\rho_{t}+p_{t}\right).
\end{equation} 
%From the expression for the apparent radius, Eq.(\ref{eq:apparent_radius}) we have $\dot{\tilde{r}}_{A} = -H\tilde{r}_{A}^{3}\left(\dot{H} - k/a^{2}\right)$. Then the above equation (\ref{eq:dynamicequation_f(R)}) can be reduced to 
Using the continuity equation (\ref{eq: Gcontiniutyeqn_non-equilibrium}) the above equation can be expressed as 
\begin{align}\label{eq:dynamiceq_fR_rA2}
-\frac{\dot{\tilde{r}}_{A}}{H\tilde{r}_{A}^{3}}+\dfrac{1}{\tilde{r}_{A}^{2}} =&  \frac{16\pi G_{eff}}{n(n-1)} \left(\rho_{t}+\rho_{\Lambda}\right)\nonumber\\&+ \frac{8\pi G_{eff}}{n(n-1)H} \dot{\rho}_{t} + \frac{8\pi\dot{G}_{eff}}{n(n-1)H} \left(\rho_{t}+\rho_{\Lambda}\right)  .
\end{align}

Once we have the expansion law as in the above Eq. (\ref{eq:dynamiceq_fR_rA2}), it is easy to derive the Friedmann equations.
 %\textbf{
 Multiply 
 %$2\dot{a} a$ on 
 both sides of the above equation with $2\dot{a} a$, we get (using $ H=\dot{a}/a $)
%}
 \begin{align}\label{eq:dynamiceq_fR_rA3}
 -\frac{2\dot{\tilde{r}}_{A}}{\tilde{r}_{A}^{3}}a^{2}+\dfrac{1}{\tilde{r}_{A}^{2}}2\dot{a} a =&  \frac{16\pi }{n(n-1)}\bigg[2\dot{a} aG_{eff} \left(\rho_{t}+\rho_{\Lambda}\right) \nonumber\\ &+ a^{2}G_{eff} \dot{\rho}_{t} + a^{2}\dot{G}_{eff} \left(\rho_{t}+\rho_{\Lambda}\right)\bigg].
 \end{align}
 The above equation can be identified as the exact differential of the form 
  \begin{equation}\label{eq:dynamiceq_fR_diff}
 \dfrac{d}{dt}\left(\frac{a^{2}}{\tilde{r}_{A}^{2}}\right) =  \frac{16\pi }{n(n-1)}\dfrac{d}{dt}\left[a^{2}G_{eff} \left(\rho_{t}+\rho_{\Lambda}\right)\right].
 \end{equation}
 On integrating the above equation we get the first Friedmann equation in f(R) gravity
\begin{equation}\label{eq:1fieldeqn_f(R)}
H^{2}+\frac{k}{a^2}  = \dfrac{16\pi G_{eff}}{n(n-1)}\left(\rho_{t}+\rho_{\Lambda}\right). 
\end{equation}
From which, one can get the second Friedmann equation by taking differential of Eq. (\ref{eq:1fieldeqn_f(R)}), using the continuity equation (\ref{eq:fRcontiniutyeqn_non-equilibrium}), 
\begin{equation}\label{eq: 2fieldeqn_f(R)}
 \dot{H} - \frac{k}{a^2} = -\dfrac{8\pi G_{eff}}{(n-1)}\left(\rho_{t} + p_{t}\right).
\end{equation}
%  deduced from the expansion law in f(R) gravity can also be made by combining the field equations (\ref{eq: 1fieldeqn_f(R)}, \ref{eq: 2fieldeqn_f(R)}).
Hence the expansion law is consistent with Friedmann equation in f(R) gravity. 

In reference \cite{Tu_2013}, the authors attempt to derive the expansion law 
%reduced from a modified expansion law 
in $f(R)$ gravity by taking $N_{sur}=4S = Af^{\prime}(R)/l_{p}^{n-1}$. 
%and without modifying 
%keeping $dV/dt.$ 
However, the resulting dynamical equation seems erroneous and hence is not consistent with the standard Friedmann equation.
 In extending this work, authors in \cite{PhysRevD.88.084019} 
 %One of the previous attempt to generalize the expansion law to the $ f(R) $ gravity in flat FRW Universe, employed the 
 %was using the 
% entropic definition of the surface $DoF$,$N_{sur}= 4S = Af^{\prime}(R)/l_{p}^{n-1}$ \cite{PhysRevD.88.084019}. 
 %W
 %There the authors 
 generalized  $\dfrac{dV}{dt}$ in the expansion law with %(\ref{eq:expansionlaw}) 
 %the change in $DoF$ on surface, 
 $\dfrac{\alpha}{(n-1)H}\dfrac{dN_{sur}}{dt}$ and proposed a dynamic equation (Eq.16 in the reference), which is valid only 
 %satisfies 
 in 
 %the 
 equilibrium conditions, 
 %condition 
 $\dot{f_{R}}=0$. Also, in the process of derivation, the authors seem to have omitted 
 %But in general the resulting dynamic equation in $ (3+1) $ FRW Universe should have an additional 
 a term, $H\dot{f}'(R)/2f(R)$, with which 
 it is impossible to arrive at the 
 %makes the system becomes 
 %and is 
 %inconsistent with the 
 Friedmann equation in $f(R)$ gravity\cite{PhysRevD.88.084019}. In contrast to these, our approach generalizes the expansion in non-equilibrium conditions and  is consistent with the Friedmann equation.
\section{Conclusions} \label{sec.5}
Padmanabhan proposed the law of expansion based on the concept of the emergence of cosmic space as cosmic time evolves. However, later this law was derived by applying the first law of thermodynamics to the horizon of expanding universe \cite{FLDezaki, Mahith2018}. In this paper, our main aim was to obtain the expansion law using the non-equilibrium first law of thermodynamics. For gravity theories having higher-order curvature corrections, the first law indeed has the non-equilibrium form due to an additional entropy generation because of the non-equilibrium evolution. Earlier, the expansion law was derived by explicitly using the corresponding Friedman equation \cite{Mahith2018}. To avoid this circular process, we first described a method to derive the expansion law by projecting the first law of thermodynamics at the horizon of an FRW Universe without explicitly using the Friedmann equations. The final form we arrived at is structurally different from one usually seen in the literature. We then extended this method to the non-equilibrium first law of thermodynamics. 

We have first formulated 
%a generic derivation for 
the expansion law as direct time evolution of the entropy, which is caused by the discrepancy between the DoF on the horizon and that within the bulk enclosed by the horizon. As a matter of fact, we have derived the expansion law from both forms of the thermodynamics law,
% \\ in (n+1) FRW Universe from the generalized first law of thermodynamics 
$-dE=TdS$ applicable to the locality of the horizon and $dE=TdS + WdV$, which is true for the entire volume of the horizon. The respective surface DoF, $N_{sur}$ is formulated as the integral over the respective entropy as given in Eq. (\ref{eq:gen surDoF}), which automatically guarantees the use of the areal volume for the correct formulation of the expansion law. Since many have shown that the use of proper invariant volume does not lead to the expansion law in any gravity theory \cite{Hareesh_2019, Chang-Young:2013gwa}. 

Further, we 
%investigated the generic 
derived the expansion law 
%in non-equilibrium 
from the modified first law in non-equilibrium, $-dE = TdS+Td_{i}S $ and $dE= TdS+Td_{i}S +WdV,$ by extending the procedure developed for the equilibrium situation. 
% We have derived the expansion law in non-equilibrium (\ref{eq:G.expansionlaw non-equilibrium}) with the degrees of freedom on the surface $N_{sur}$ and that on the bulk $N_{bulk}$ in Eq.(\ref{eq: gen DoF_non-equilibrium}). The generic derivation shows that the entropy evolution (or the emergence of areal volume) is proportional to the discrepancy in the degrees of freedom on the surface and in bulk,$N_{sur}-\epsilon N_{bulk}$  even in the non-equilibrium situation, where the effective change in entropy has an additional entropy production term $d\bar{S}=dS + d_{i}S$. 
Then we particularity formulated the expansion law in f(R) gravity using the generalized expansion law in non-equilibrium (\ref{eq:G.expansionlaw non-equilibrium}). We found that %obtained 
%the 
the expansion law obtained for f(R) gravity (\ref{eq:f(R)_expansionlaw non-equilibrium}) is consistent with the Friedmann equations in f(R) gravity. This shows that the adequate entropy evolution (or the emergence of areal volume) in f(R) gravity is proportional to $N_{sur}-\epsilon N_{bulk}$, the discrepancy in the $DoF$.

It is to be noted that 
%On 
the generic derivation of the expansion law in non-equilibrium, we used the unified formulation for modified gravities with an effective dynamic coupling strength $G_{eff}$ \cite{PhysRevD.90.104042} to keep the generality of the derivation. Hence the resulting expansion law (\ref{eq:G.expansionlaw non-equilibrium}) can be widely applicable to several minimally coupled modified gravity theories like f(R), generalized Brans-Dicke, scalar-tensor-chameleon and $ f(R,\mathcal{G}) $ generalized Gauss-Bonnet gravity.  Thus, in general, we can conclude that there exists a strong correlation between generalized first law and the expansion law, both in equilibrium and non-equilibrium thermodynamic conditions.  %which is in terms of degrees of freedom; one imply the other or vies-versa. 

\acknowledgments
 
 Hassan Basari V.T acknowledges Cochin University of Science and Technology for financial support. P. B. Krishna acknowledges KSCSTE, Govt. of Kerala for the fellowship.

%\bibliographystyle{apsrev4-1}
%\bibliography{ref}
%
\end{document}